\pdfoutput=1
\documentclass[journal]{IEEEtran}

\usepackage[nolist,nohyperlinks]{acronym}
\usepackage{mathtools}
\usepackage{breqn}
\usepackage{amsmath}
\usepackage{balance}
\usepackage{cite}
\usepackage{framed}
\usepackage{siunitx}
\usepackage{soul}
\usepackage{multirow}
\usepackage{etoolbox}
\usepackage{verbatim}
\usepackage{graphicx}
\usepackage{subcaption}
\usepackage{color}
\usepackage[table,xcdraw]{xcolor}
\usepackage{float}
\graphicspath{{./images/}}

\usepackage{url}

\newtoggle{authorcomment}
\toggletrue{authorcomment}

\usepackage{algorithm}
\usepackage{algorithmic}

\newcommand{\squeezeup}{\vspace{-5mm}}

\begin{document}
\bstctlcite{IEEEexample:BSTcontrol}

\title{Mobility for Cellular-Connected UAVs: Challenges for the network provider}

\author{
  \IEEEauthorblockN{
    Erika Fonseca*,
    Boris Galkin*,
 Marvin Kelly$^\dagger$,
    Luiz A. DaSilva$^\ddag$,
    Ivana Dusparic*
  }\\
  \IEEEauthorblockA{
    \IEEEauthorrefmark{1} CONNECT - Trinity College Dublin, Ireland,\\
     $^\dagger$ Dense Air Ltd.,\\
    $^\ddag$Commonwealth Cyber Initiative, Virginia Tech, USA \\
    \{fonsecae, galkinb, duspari\}@tcd.ie,  ldasilva@vt.edu}\\
    
\textit{“This work has been submitted to the IEEE for possible publication. Copyright may be transferred without notice, after which this version may no longer be accessible.”}
  }

\maketitle

\begin{acronym}[IMT-Advanced]
  \acro{3gpp}[3GPP]{3\textsuperscript{rd} Generation Partnership Program}
\acro{5g}[5G]{Fifth Generation Mobile Networks}
\acro{Adam}{Adaptive Moment Optimisation}
\acro{ap}[AP]{Access Point}
\acro{anr}[ANR]{Automatic Neighbouring Relation}
\acro{amf}[AMF]{Access and Mobility Management Function}
\acro{atc}[ATC]{Air Traffic Control}
\acro{bs}[BS]{Base Station}
\acro{bler}[BLER]{Block Error Rate}
\acro{cc}[C\&C]{Command \& Control}
\acro{cdf}[CDF]{cumulative distribution function}
\acro{cdma}[CDMA]{Code Division Multiple Access}
\acro{cdr}[CD\&R]{Conflict Detection \& Resolution}
\acro{cfo}[CFO]{Carrier Frequency Offset}
\acro{comreg}[ComReg]{Commission for Communications Regulation}
\acro{csi}[CSI]{Channel State Information}
\acro{easa}[EASA]{European Aviation Safety Agency}
\acro{dql}[DQN]{Deep Q-Learning}
\acro{esn}[ESN]{echo state network}
\acro{ecgi}[ECGI]{evolved cell global identifier}
\acro{faa}[FAA]{Federal Aviation Administration}
\acro{fso}[FSO]{Free-Space Optical}
\acro{geo}[GEO]{Geosynchronous Equatorial Orbit}
\acro{gps}[GPS]{Global Positioning System}
\acro{gs}[GS]{Ground Station}
\acro{gnb}[gNB]{next generation NB}
\acro{gue}[GUE]{Ground User Equipment}
\acro{hap}[HAP]{High Altitude Platform}
\acro{ho}[HO]{Handover}
\acro{hof}[HOF]{Handover Failure}
\acro{iot}[IoT]{Internet of Things}
\acro{kpi}[KPI]{Key Performance Indicator}
\acro{lap}[LAP]{Low Altitude Platform}
\acro{leo}[LEO]{Low Earth Orbit}
\acro{los}[LoS]{Line-of-Sight}
\acro{lte}[LTE]{Long Term Evolution}
\acro{mac}[MAC]{Media Access Control}
\acro{mcp}[MCP]{Matern Cluster Process}
\acro{mc}[MC]{Monte Carlo}
\acro{mimo}[MIMO]{Multiple Input Multiple Output}
\acro{mip}[MIP]{Mixed-Integer Programming}
\acro{milp}[MILP]{Mixed-Integer Linear Programming}
\acro{mm}[MM]{Mapping Mechanism}
\acro{ml}[ML]{machine learning}
\acro{mno}[MNO]{Mobile Network Operator}
\acro{nlos}[NLoS]{non-Line-of-Sight}
\acro{nn}[NN]{Neural Network}
\acro{nrt}[NRT]{Neighbour Relation Table}
\acro{ofdma}[OFDMA]{Orthogonal Frequency Division Multiple Access}
\acro{oam}[OAM]{Operations Administration and Maintenance}
\acro{osi}[OSI]{Open Systems Interconnection}
\acro{otdoa}[OTDoA]{Observed Time Difference of Arrival}
\acro{ott}[OTT]{Over-The-Top}
\acro{pv}[PV]{photo-voltaic}
\acro{pdf}[pdf]{probability density function}
\acro{ppp}[PPP]{Poisson Point Process}
\acro{pci}[PCI]{physical cell identifier}
\acro{qos}[QoS]{Quality of Service}
\acro{rat}[RAT]{Radio Access Technology}
\acro{rc}[RC]{Remote Control}
\acro{rl}[RL]{Reinforcement Learning}
\acro{nr}[NR]{Neighbouring Relation}
\acro{rss}[RSS]{Received Signal Strength}
\acro{rrc}[RRC]{Radio Resource
Control}
\acro{rsrp}[RSRP]{Reference Signal Received Power}
\acro{se}[SE]{Spectral Efficiency}
\acro{sir}[SIR]{Signal-to-Interference Ratio}
\acro{sinr}[SINR]{Signal-to-Interference-and-Noise Ratio}
\acro{snr}[SNR]{Signal-to-Noise Ratio}
\acro{son}[SON]{Self-Organising Network}
\acro{uav}[UAV]{Unmanned Aerial Vehicle}
\acro{ulid}[ULID]{Uplink ID}
\acro{ue}[UE]{User Equipment}
\acro{ula}[ULA]{Uniform Linear Array}
\acro{wsn}[WSN]{Wireless Sensor Network}
\acro{wsn}[WSN]{Wireless Sensor Network}
\acro{rv}[RV]{random variable}
\acro{ppp}[PPP]{Poisson point process}
\acro{pgfl}[PGFL]{point generation functional}
\acro{pdf}[PDF]{probability density function}
\acro{pso}[PSO]{particle swarm optimisation}
\acro{3G}{Third Generation Mobile Networks}

\end{acronym}
\begin{abstract}
 \ac{uav} technology is becoming more prevalent and more diverse in its application. 
 5G and beyond networks must enable \ac{uav} connectivity. This will require the network operator to consider this new type of  user in the planning and operation of the network. 
 This work presents the challenges an operator will encounter and should consider in the future as \acp{uav} become users of the network. We analyse the 3GPP specifications, the existing research literature, and a publicly available \ac{uav} connectivity dataset, to describe the challenges.
We classify these challenges into network planning and network optimisation categories. We discuss the challenge of planning network coverage when considering coverage for flying users and the PCI collision and confusion issues that can be aggravated by these users. In discussing network optimisation challenges, we introduce \ac{anr} and handover challenges, specifically the number of neighbours in the \ac{nrt}, and their potential deletion and block-listing, the frequent number of handovers and the possibility that the \ac{uav} disconnects because of handover issues.  
We discuss possible approaches to address the presented challenges and use a real-world dataset to support our findings about these challenges and their importance. 
\end{abstract}


\IEEEpeerreviewmaketitle

\acresetall

\section{Introduction}
\label{sec:intro}

\acp{uav} are expected to make use of \ac{5g} connectivity when performing building inspections (roofs, chimneys, siding), security surveillance, search and rescue operations, mapping, agricultural surveys, delivery of goods, live streaming of shows and events, etc \cite{examples}. Although the regulatory bodies have not yet defined how this integration will happen, \ac{uav} connectivity is the focus of a number of research efforts in the \ac{3gpp} \cite{release14,3GPP_2018,3GPP_2020}.
\ac{3gpp} Release 14~\cite{release14}, for example,  states that a UAV needs to maintain continuous connectivity with the cellular  network while flying at speeds of up to 300 km/h.

Qualcomm has carried out several experiments to analyse the viability of using the existing cellular  network for providing connectivity to \acp{uav} \cite{qualcom}. They report that the \acp{uav} can have a connection to the cellular  network through the side lobes of the antennas that have their main lobe pointed to the ground, where the \ac{gue} typically operates.
Preliminary results show that the coverage is adequate for \acp{uav} flying up to 120 m above ground \cite{qualcom}. 
Obstacles between \ac{ue} and the cells can deteriorate the signal. The \ac{uav} coverage is adequate because, with the greater height of the \acp{uav}, there are no obstacles between them and the antennas. 
However, at greater heights, the increased \ac{los} to multiple cells results in high levels of interference at the \acp{uav}, which poses handover and mobility management challenges~\cite{mohamed2018memory}.

Handover is the process by which a \ac{ue} changes its serving cell. It is typically triggered when the \ac{ue} moves out of the coverage area of its current serving cell. Ideally, the handover should be seamless to the \ac{ue}, such that it would not suffer any data interruption during the process. If a \ac{ue} experiences multiple handovers, a handover delay might occur, resulting in substantial deterioration to the \ac{ue} \ac{qos} \cite{pingpong}.
To proceed with a handover, the \ac{ue} needs to detect pilot signals from neighboring cells. The list of neighbour cells is defined on the \ac{nrt} that is stored in the connected cell. 
In \ac{5g} this list is generated locally  by \ac{anr}, based on \ac{ue} measurements of  \ac{rsrp} from nearby cells. \ac{anr} was introduced in \ac{3G} and  was shown to reduce planning and operational costs for operators \cite{anrbetter}. With the more stringent latency and data rate requirements in 5G,  seamless handovers are even more critical, which places a greater importance on the efficient use of \ac{anr}.

Unlike \acp{gue}, \acp{uav} will sense a large number of cells \cite{qualcom}, which leads to a considerable increase in the size of the \ac{anr} at the serving cell and  increase  the complexity of handover decisions~\cite{why-not-long-list}. 

This paper comprises the following sections: 
In Section \ref{sec:anr}, we review mobility management in 5G networks. In Section \ref{sec:uav}, we identify challenges that \ac{uav} mobility can bring to future networks and propose approaches to mitigate these challenges. We illustrate the challenges using a publicly available available dataset of measurements taken by an \ac{uav} flying in the city of Dublin, and connecting to a two-tier network. In Section \ref{sec:conclusion}, we conclude the paper by discussing directions for future work.

\section{Mobility management in 5G}
\label{sec:anr}

\begin{figure}
  \centering
  \includegraphics[width=0.75\linewidth]{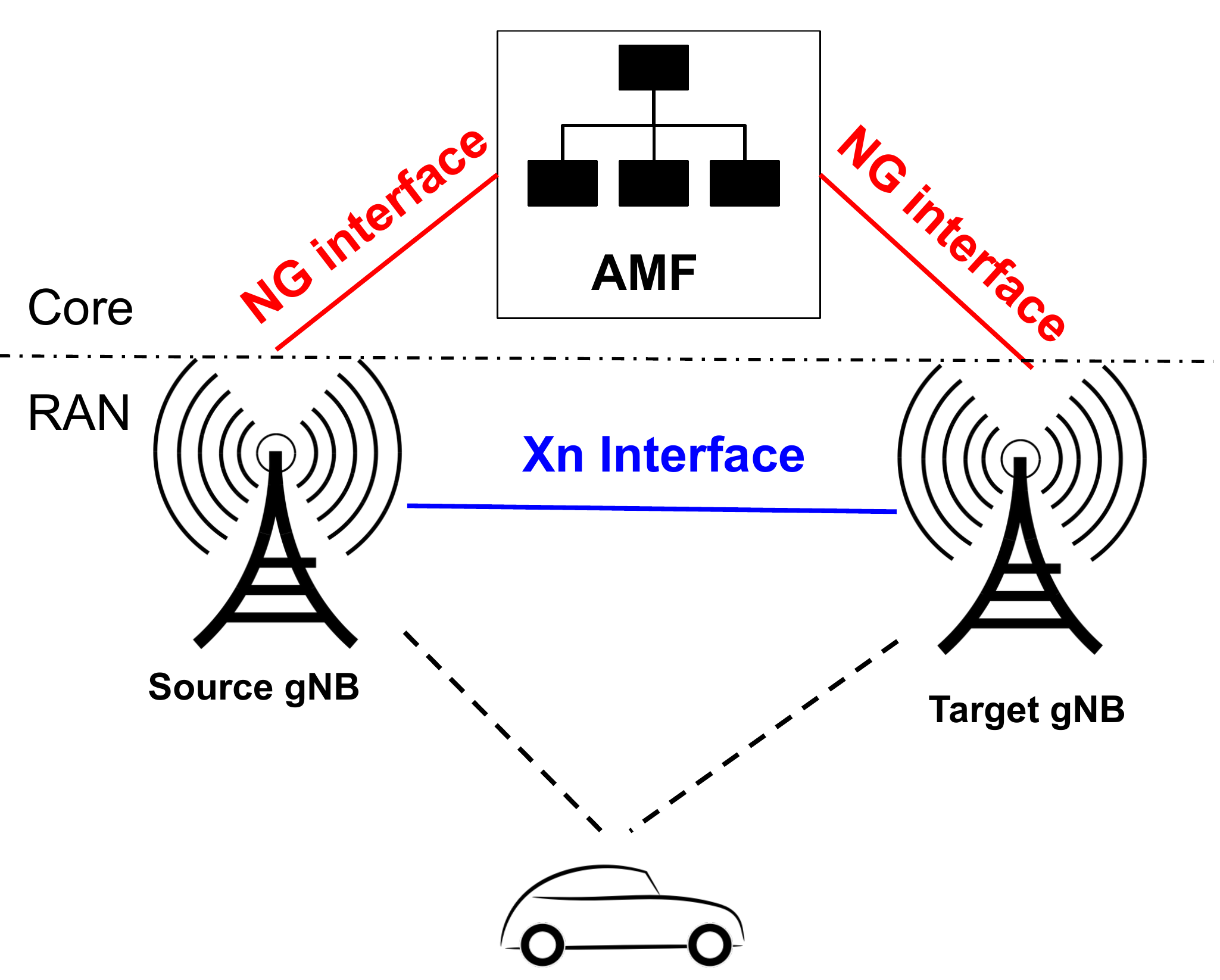}
    \caption{Entities involved in a \ac{ue} handover.}
    \label{fig:arch}
    \vspace{-4.5mm}
\end{figure}

Mobility management in 5G is performed by three main entities, illustrated in Figure \ref{fig:arch} and specified in \cite{3GPP_331}.
These are the \ac{amf}, \ac{gnb} (that is the \ac{bs} equivalent as defined in 5G, which may comprise one or more cells)\footnote{In this paper, we refer to the serving BS as \ac{gnb} as we consider the \ac{uav} connected to the 5G network. When we use the generic term \ac{bs} we refer to any technology \ac{bs}, not necessarily 5G.}, and the \ac{ue}. \ac{amf} is responsible for handling connection and mobility management for \acp{ue}. The \ac{gnb} provides connection to the \ac{ue}; it has a connection to \ac{amf} via the NG interface and to other \acp{gnb} via the Xn interface. The last entity is the \ac{ue} itself. 

In 2G and 3G networks, the \ac{nrt} is deployed as part of the operations and maintenance system, which is equivalent to OAM (Operations Administration and Maintenance) in 5G. In 5G the \ac{gnb} has the permission to create new entries in the \ac{nrt}. The \ac{anr} determines which cell should be added based on \ac{ue} measurements and OAM updates. 
The \ac{ue} can perform measurements to check for new cells, measure signal quality, determine if it needs to make a handover, or add a new cell to the \ac{anr} \cite{3GPP_331}.

The purpose of this procedure is to transfer measurement results from the \ac{ue} to the network in order to allow the network to decide how to improve performance for the \acp{ue} and the network itself. The \ac{ue} can initiate the measurements only after successful security activation in the network.

These measurements occur as often as determined by the \ac{gnb} and vary based on the implementation of each operator. 
If the measurement is made in the same frequency band (intra-frequency) it can be done without any specific preparations to make the measurements. If the measurements are in another frequency (inter-frequency) the network needs to schedule a measurement gap where the \ac{ue} stops receiving and transmitting data, changes to the frequency where it has to make the measurements, and senses it in order to find more suitable \ac{bs}s. These gaps can affect the performance observed by the \ac{ue} if the \ac{ue} is in dedicated mode (transmitting and receiving data). In idle mode, the \ac{ue} can perform the inter-frequency measurements without impacting its \ac{qos}. 
The measurements are sent to the serving cell, which uses them to check for events to trigger a handover, or to add a new cell to the \ac{anr}, for example. 


The information regularly decoded from a  measurement by the \ac{ue} includes the local identifier of the cell, named \ac{pci} in LTE and 5G. If the \ac{pci} is not in the \ac{nrt}, then the serving cell can send a message instructing the \ac{ue} to sense the \ac{ecgi} of that cell, that is its global ID, in order to introduce this new cell into the \ac{anr}. 
To determine the \ac{ecgi}, the \ac{ue} needs to decode more data from the sensed \ac{bs}, and to decode the \ac{ecgi} it needs more than a single measurement gap. If the \ac{ue} is in connected mode, actively receiving and transmitting data, the \ac{ue} might not have time to perform the inter-frequency measurement and to decode the \ac{ecgi}, as a result of which the \ac{ue} might be disconnected.

The  mobility events defined by the 3GPP \cite{3GPP_331} that can happen after the measurements are made and passed to the \ac{gnb} are described below. They are divided into intra-\ac{rat}, denoted as events A, and inter-\ac{rat}, denoted as events B.

\begin{itemize}
    \item {Event A1}: The serving  cell  signal  becomes  better  than  an operator-defined  signal  quality  threshold, i.e. the cell is providing  good  signal  quality. This event is commonly used to cancel an ongoing handover procedure, to avoid a ping-pong effect from the handover.
    \item {Event A2}: The serving cell signal becomes worse than an operator-defined signal quality threshold, i.e. the cell is not providing a good signal quality. 
    This event can trigger  Inter-RAT measurements, for example, as new connectivity options must be considered for the \ac{ue}.
    \item {Event A3}: The neighbour cell signal becomes better than the serving cell signal by a certain offset. This event can trigger the handover process to the neighbour cell.
    \item {Event A4}: The neighbour cell signal becomes better than an operator-defined signal quality threshold. This event is commonly used to trigger a handover. In this event, the handover is not triggered by the radio-signal conditions, but due to a network strategy specified by the operator, such as load balancing across cells, for example.  
    
    \item {Event A5}: The operator defines 2 thresholds, refereed to as threshold1 (with lower value) and threshold2 (with higher value) in 3GPP. This event occurs when the serving cell signal becomes lower than threshold1 and the neighbour cell signal higher  than threshold2. This event can trigger a handover based on the absolute measured signal strength values. This time-critical handover can be useful if the \ac{ue} is leaving the serving cell coverage area and needs to handover, even if the target cell is not better by an offset than the serving cell to trigger an event A3.
    \item {Event A6}: The neighbour cell signal becomes higher by an offset than the serving secondary cell signal. In the case the \ac{ue} has a multi-connection to more than one \ac{bs}, and it can trigger a handover from  its current secondary cell to a new one.
    \item {Event B1}: An inter-\ac{rat} neighbour provides a stronger signal than an operator-defined signal quality threshold. This event may trigger a inter-\ac{rat} handover.
    \item {Event B2}: The operator defines 2 thresholds, referred to as threshold1 (with lower value) and threshold2 (with higher value). The signal from the serving cell becomes lower than threshold1, and an inter-RAT neighbour provides a signal higher than threshold2. This event can trigger an inter-RAT handover.
\end{itemize}

Although the \ac{ue} measurements are already described in the 5G standard \cite{3GPP_331}, in LTE there are two other ways in which a cell may be added to an \ac{anr} \cite{ramachandra2016automatic}, which might be adopted in 5G depending on the individual operator. The first alternative to the \ac{ue} taking measurements is the \ac{ue} transmitting an Uplink ID, which should be unique locally. The cells that detect the signal above a certain threshold will add the serving cell of the source \ac{ue} into their \ac{nrt}. Another possible solution is to add a cell to the table once a \ac{ue} loses connection and re-connects in a new cell. The new cell would add the last cell to which the \ac{ue} was connected into its \ac{nrt}. As this method makes use of a \ac{ue} disconnect, it cannot be applied if the operator wants to provide seamless handover at all times.

\section{UAV Mobility Challenges}
\label{sec:uav}

\acp{uav} were introduced as a new type of user of the cellular  network in \ac{lte} and they are expected to increase in numbers and applications in 5G networks and beyond. The requirements in \ac{3gpp} release 15 for \ac{uav} connectivity to the network are summarized in Table \ref{table:requirements}. To meet these requirements and the even stricter requirements in future releases of 5G, the network will need to adapt to able to serve the connected \acp{uav}. One of the biggest challenges for connected \acp{uav} is the presence of simultaneous \ac{los} channels with several cells which may be far away. In \cite{qualcom}, authors demonstrated via simulations and experiments that a \ac{uav} can sense significantly more cells than a \ac{gue}. 
The fact that \acp{uav} can detect  a larger number of cells across a greater area means that the network should treat the \ac{uav} \ac{ue} differently from a \ac{gue}, in terms of mobility management.
The principal mobility-related challenges that a \ac{uav} can introduce to 5G networks operators are discussed in this section.

\begin{table}[h]
\centering
\begin{tabular}{|c|c|}
\hline
\rowcolor[HTML]{C0C0C0} 
\multicolumn{1}{|c|}{\cellcolor[HTML]{C0C0C0}\textbf{Parameters}} & \multicolumn{1}{c|}{\cellcolor[HTML]{C0C0C0}\textbf{Value}} \\ \hline
Latency for traffic & 50ms \\ \hline
UL/DL data rate & 200kbps \\ \hline
Application data rate (UL) & up to 50 Mbps \\ \hline
UAV \ac{ue} height & up to 300 m \\ \hline
UAV \ac{ue} velocity & up to 160 km/h \\ \hline
\end{tabular}
\caption{
UAV requirements in 3GPP Release 15.
\squeezeup
	}
	\label{table:requirements}
\end{table}

\subsection{Network Planning challenges}
Before the cellular  network starts its operation,  the operators need to plan the geographic locations of the \acp{gnb}, along with configuration parameters such as their antenna azimuth and mechanical tilt. If \acp{uav} become a significant user of the network, they  need to be taken into consideration from the planning stage of  network deployment. This section discusses  challenges encountered at the planning stage. 

\subsection*{Network coverage planning}
Network coverage planning is essential to avoid interference and unnecessary handovers. For the previous generations of cellular  networks, the coverage was planned only for \acp{gue}, and the main lobe of the \ac{bs} antennas was often the only one taken into account. For the next generations of cellular  network, the coverage needs to be planned to also include \ac{uav} \acp{ue}, and needs to consider what kind of network coverage will be provided in the air. 
A common way to plan a cellular network is by using software tools that consider 3D maps of a given area and antenna radiation patterns. 
To integrate \ac{uav} users, the tools used to plan the network coverage need to be adapted to consider antenna side lobes and should also project the signal propagation into the sky. 

Another critical part of  network planning that becomes harder with the introduction of a flying \ac{ue} is the PCI distribution. The flying \ac{ue}s can exacerbate  PCI confusion and collision, which have been reported in LTE networks and persist for 5G networks. 
Usually, the \ac{pci} planning is made to allocate concurrent \acp{pci} to \acp{bs} that are distant from each other, to ensure that a \ac{ue} will be unlikely to detect the same \ac{pci} being transmitted by more than one \ac{bs} at a time. However, considering connected \acp{uav}, it will be necessary to understand the air coverage in advance to plan the \ac{pci} distribution. Next, we discuss PCI confusion and collision challenge and why \ac{uav} users aggravate it.

\subsection*{PCI challenges}
In Section \ref{sec:anr}, we introduced the events triggered after the measurement reports. The first piece of information a \ac{ue} senses about a neighbouring \ac{gnb} is the \ac{pci}, that is the local cell identifier. Each cell in 5G or LTE has its own PCI.
If the PCI assignment is poorly planned, it can affect the handover process and delay the downlink synchronisation. Another possible consequence is  increased \ac{bler} and decoding failures of physical channels.
In LTE, there are 504 unique PCIs, compared to 1008 in 5G. If there are different tiers of the network, the network needs to divide the PCIs for each tier. 

Consider a two tier network with macro-cells and small-cells, for example. The PCI values contained in set A will be reserved to the macro-cells and those in set B for small-cells. A and B have no intersection. This rule cannot be violated inside the same network. This division decreases the number of possible PCIs for each tier, which can aggravate the issue of PCI availability. 
Due to the fact that the \acp{gue} usually connect to cells that are close to them, with good network planning it is possible to avoid most cases of PCI collision and confusion for \acp{gue}.

Figure \ref{fig:pci} illustrates a well-planned network, where concurrent PCIs have a significant distance between them, which means that PCI confusion is not likely to happen for \acp{gue}. The main issue occurs when a \ac{uav} flies overhead, as it senses more distant cells that can have the same PCI as the serving cell, which results in PCI collision, or be already on the \ac{nrt} of the serving cell, which results in the PCI confusion. Both issues are detailed below.

\subsubsection*{PCI Confusion}
PCI confusion happens when the detected \ac{pci} is in the \ac{nrt} of the serving cell. The serving cell assumes that the sensed cell is already in the \ac{nrt} and does not request a check of the \ac{ecgi}.
The situation is made worse in the scenario where the \ac{uav} tries to handover to this concurrent cell because all of the handover configuration will be carried out with the wrong cell and the \ac{ue} could have its connection broken. The opposite can also happen: if a \ac{uav} adds a distant cell to the list and a \ac{gue} senses a closer cell with the same \ac{pci} the closer one would not be added to the \ac{nrt}, which would result in the handover configuration being sent to the far away cell.
It may even result in the concurrent cell being added to a block-list, as many attempts to handover to this cell would fail.  A neighbour should be block-listed if there are repeated attempts of unnecessary connections, and once block-listed, the cell is not an option for handover anymore. 

\begin{figure}
  \centering
  \includegraphics[width=0.8\linewidth]{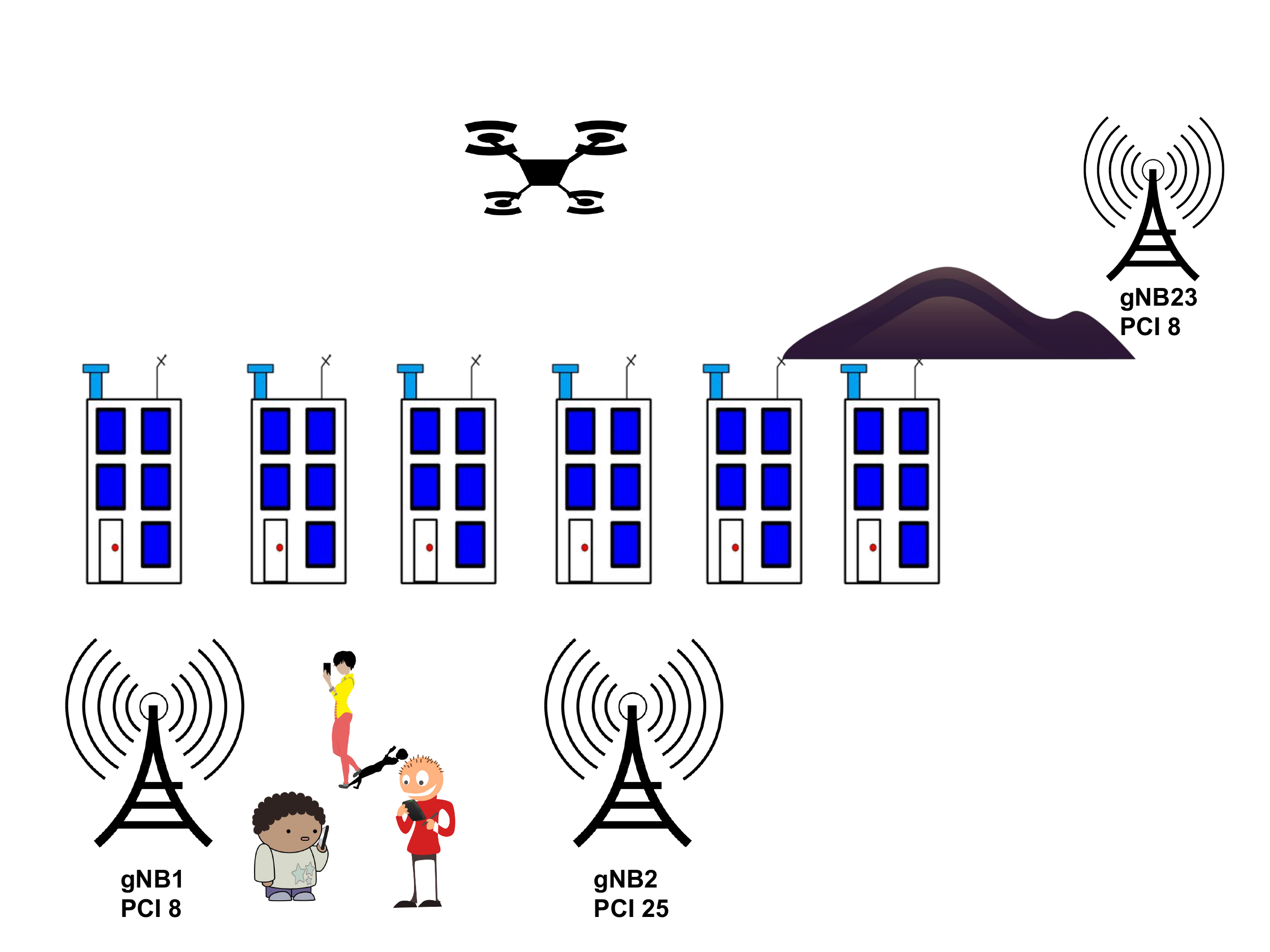}
    \caption{PCI confusion/collision challenge.}
    \label{fig:pci}
    \vspace{-3.5mm}
\end{figure}

As an example, assume that in Figure \ref{fig:pci} the \ac{uav} is connected to the \textit{\ac{gnb}2}. In the \textit{\ac{gnb}2} NRT, the \textit{\ac{gnb}1} is a neighbour, and its PCI is saved in the table corresponding to \textit{\ac{gnb}1}. Once the UAV flies and senses a strong signal from \textit{\ac{gnb}23}, it detects its PCI. As the PCI of \textit{\ac{gnb}23} is the same as that of \textit{\ac{gnb}1}, the serving \ac{gnb}, \textit{\ac{gnb}2}, decides that the signal sensed by the UAV is from \textit{\ac{gnb}1} and does not ask the UAV to verify the ECGI. If the UAV tries to handover to \textit{\ac{gnb}23}, all of the configuration for handover will be sent to \textit{\ac{gnb}1}, and the network might not be able to detect that there is a problem before the UAV disconnects.

\subsubsection*{PCI collision}
PCI collision happens when two cells that cover the same area are allocated with the same PCI. In this situation, the \ac{ue} connected to one of them will not sense for another cell with the same PCI, which can result in the \ac{ue} not being connected to the best serving cell. For example, consider that the UAV is going in the direction of the hill and is connected to \textit{\ac{gnb}1}. Even if \textit{\ac{gnb}23} has a strong signal and is the only \ac{gnb} available in that direction, the UAV will not consider it as as option and will disconnect before trying to connect to \textit{\ac{gnb}23}.

A possible consequence of PCI confusion and collision is that the network has to be updated with more appropriate PCIs once these issues happen. To update the PCI of a cell, the \ac{gnb} needs to be restarted, which can take more than one hour.

To solve the PCI distribution issue, one possible solution would be for UAVs to have two radios for communication and measurements. Radio one (R1), would be used for communication, but its priority would be sensing. Radio two (R2), would be used for communication only. When the UAV needs to sense and make measurements, we propose that the UAV would always sense the \ac{ecgi} directly to avoid PCI confusion/collision. During the measurements, R1 should stop any communication that could be using the radio. R2 would not stop its transmission and data reception at any time during the measurement reports. This method would ensure that UAV does not lose connection during the measurements. The drawback of this approach is that having two radios is more expensive and takes up additional space on the device. 
Nevertheless, the use of two radios should be considered by vendors and regulators.

To support our earlier claim that the UAV \ac{ue} should be taken into account by network operators during the various steps of  network planning, 
we made use of the dataset available in \cite{satori} with signal to noise power measurements made by a \ac{uav}-mounted handset. The network is a two-tier cellular  network in Dublin city centre that operates in the 3.6GHz band. The discussion in this paper focuses on the small-cell measurements. 

\begin{figure}
  \centering
  \includegraphics[width=0.6\linewidth]{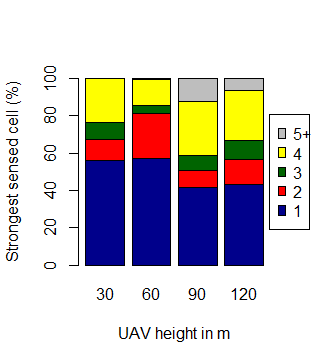}
    \caption{Percentage of time the UAV was sensing the n-th closest cell as the strongest signal cell.}
    \label{fig:closest}
    \vspace{-3.5mm}
\end{figure}

Typically, for \acp{gue} it is a fair assumption that the \ac{ue} will be connected to the closest cell, a common assumption made by the research community \cite{erika-height,rami3}. This analysis investigates how often the UAV sensed the strongest signal as coming from the geographically closest cell during its flight. Figure \ref{fig:closest} illustrates the most potent sensed cell relative to its distance to the UAV, for four different altitudes, 30m, 60m, 90m, and 120m. At 30m and 60m, the UAV senses more than 50\% of the time the strongest signal as coming from closest cell. The same does not happen at higher altitudes: when the UAV is at 90m and 120m, it senses the closest cell as the strongest for around 40\% of the time; for almost 30\% of the time, it senses the signal from the fourth closest cell as being the strongest one.  
The behaviour presented in the results clearly differs from the expected behaviour from a \ac{gue}.  

Figure \ref{fig:closest} reinforces the idea that the coverage in the air needs to be considered before deploying new \ac{gnb}s, as the UAVs 
can connect to much more distant \ac{gnb}s. It also highlights that the research community's assumptions that the \ac{ue} will connect to its closest \ac{gnb} is no longer holds in the case of \acp{uav}. 

\subsection{Network optimisation challenges}
Once the cellular  network is deployed, there is still the need to further optimise the network due to changes in the environment or traffic load, or to increase its performance. \ac{son}  is an automation approach introduced in LTE that is also available in 5G, designed to make the planning, configuration, management, optimisation, and error-correction of the cellular radio access network more straightforward and rapid. Once a \ac{gnb} turns on, \ac{son} will automatically configure its PCI, transmission frequency and power \cite{release8}. Our analysis into the \ac{uav} behaviour and the 3GPP specifications  show that \ac{son} should be adapted to optimise the network to accommodate \ac{uav} users. In this section we introduce the optimisation challenges that the \ac{uav} as a user can bring to the network operators. 

\subsection*{ANR challenges}
ANR is a critical feature to deal with \ac{ue} mobility management and therefore it also needs to be adapted to the new reality of the \ac{uav} network users. In this section, we  present the challenges that should be addressed for the ANR to allow the network to meet the service requirements of \acp{uav}.

\subsubsection*{Number of neighbours in \ac{nrt}}
Authors in \cite{ericssonttt} report that a \ac{uav} senses more cells compared to \acp{gue}, which can result in an increase in the number of neighbours in the \ac{nrt}. This can be detrimental to all of the \acp{ue} connected to that cell, as a \ac{ue} usually needs to sense all the cells in the neighbour table before performing a handover \cite{why-not-long-list}. If the list is too long and the \ac{ue} moves fast, the \ac{ue} might not have time to sense all the cells in the \ac{nrt} and may lose connection before performing the handover. The need to sense an excessive number of cells also goes against the ultra-lean principle, whereby the network is designed to significantly improve energy efficiency and avoid unnecessary measurements \cite{ramachandra2016automatic}. 

\subsubsection*{Block-listing neighbours} 
Once a far away neighbour is added by the \ac{uav} into the \ac{nrt}, there is a small chance that this cell will be sensed by a \ac{gue}. If it frequently happens that a \ac{gue} cannot sense the far cell, this cell will be deleted from the \ac{nrt} frequently. Depending on the Neighbour Removal Function's implementation, that can also result in this neighbour cell being added to the block-list of the serving cell. A cell can be block-listed if it is being removed frequently from the \ac{nrt}. A block-listed cell is not an option for handover for any \ac{ue} in the serving cell.
If the cell is block-listed, \acp{uav} that could benefit from a handover to that cell will no longer have this possibility.

A possible solution to the ANR problems presented above might be having a separate \ac{nrt} for flying users. This would ensure they do not interfere with the \ac{gue} connectivity and vice-versa, and it would not deteriorate their service. It would  allow network operators to design a more fine-tuned solution to the \ac{nrt} for flying users, which is a subject that has not been explored by the research community.

\subsection*{Handover challenges}
Once the PCI confusion and collision issues are resolved, additional challenges related to handover need to be addressed to ensure that UAV UEs do not overload the network and do not unnecessarily disconnect. We discuss those below. 

\subsubsection*{Frequency of handovers for \acp{uav}}
Authors in \cite{handover-experiment} reported that \acp{uav} perform, on average, five times more handovers when compared to a \ac{gue}. 
These values show that the mobility of a \ac{uav} tends to generate more signalling overhead in the network and that the parameters used to trigger Event A3 need to be adjusted for UAVs. 

\subsubsection*{Connection interruption time}
Authors in \cite{ericssonttt} show that sometimes the handover does not start for UAV users because the \ac{rsrp} measured by the \ac{uav} from neighbouring cells does not have a minimum difference of 3dB between the serving cell and the possible  handover target cell. As a result, the UAV \ac{ue} does not send event A3, which is required to trigger the handover. A consequence of this is that  \acp{uav} will experience more frequent disconnection from the network than \acp{gue}~ \cite{ericssonttt}. Once the \ac{uav} moves, it moves between side lobes and antennas nulls quickly, and there is no time to make a seamless handover, resulting in disconnection when the UAV enters the nulls of the antenna \cite{rami2}.

This indicates that the network parameters to start event A3 in \ac{gue} are not suitable for \acp{uav}. The event can trigger handovers when they are not needed, resulting in a ping-pong effect, or  not trigger handovers at the right time, resulting in disconnection. It is, therefore, necessary to introduce an adaptive threshold to start event A3 for UAVs. The threshold needs to be designed for this type of user and needs to take into account the changes in the environment to adapt quickly to the new situation.

\begin{figure}
  \centering
  \includegraphics[width=0.9\linewidth]{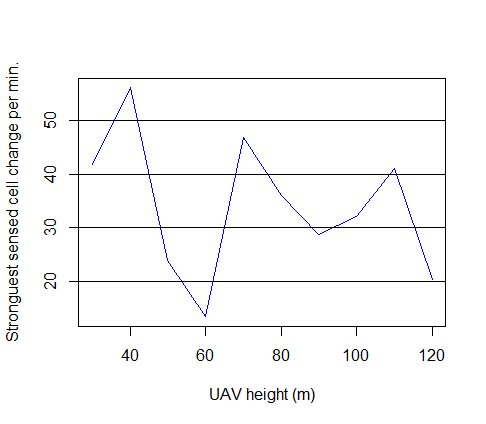}
    \caption{Number of strongest cells sensed per minute during the path per altitude.}
    \label{fig:pingpong}
    \vspace{-4.0mm}
\end{figure}

Using the data provided in \cite{satori}, we carry out an additional analysis of the small-cell deployment, by looking at how often there is a change of the strongest cell when the UAV is flying through the network. Figure \ref{fig:pingpong} illustrates how often the UAV experiences a change of strongest cell during its path. The collected values show that the strongest cell fluctuates dramatically across different heights. This is due mainly to the planned service area for the network being primarily at ground level. At other heights there are no dominant cells and hence several cells are received with similar signal levels. Further investigation is needed on how the handover performance can be optimised at these heights.

\section{Conclusion}
\label{sec:conclusion}
This paper presented and examined the main challenges that network operators may encounter when \acp{uav} become common users of the network, and proposed directions to solve some of these challenges. 
We divided the challenges into network planning and network optimisation challenges. To support our claims regarding the challenges we analysed data from a publicly available dataset which contained measurements from a UAV user connecting to a small cell network in an urban environment.

We presented the new coverage planning challenges when considering \ac{uav} \acp{ue}. Existing network tools used for coverage planning are focused on \acp{gue} and do not project how the coverage from the antenna main and side-lobes projects into the air. These tools need to be adapted to consider air coverage; to achieve this, it will be necessary to run air drive-tests to access the coverage for \acp{uav}. 
During the network planning, it is also vital to consider \acp{uav} when designing the PCI assignment, to avoid PCI collision and confusion. The typical strategy to avoid collision and confusion is to allocate concurrent PCIs to cells as distant to each other as possible. However, as \acp{uav} can sense far away cells, this might not be sufficient to avoid the PCI collision/confusion problem. We propose the implementation of two radios on the \ac{uav}, where one would prioritise sensing \acp{ecgi}, which would avoid the mentioned issues.

We also presented challenges that can occur during the optimisation of the network. The ones related to the \ac{anr} concern the large number of neighbours in the list, and the block-listing of cells. We suggested a possible direction to solve this issue by implementing a separate \ac{nrt} for \acp{uav}.
The presented challenges with the handover process included the greater number of handovers for \acp{uav}, compared to \acp{gue}, and the connection interruption time which \acp{uav} might experience due to flying into the nulls of the \ac{gnb} antennas. As a possible solution to both handover challenges, we proposed an adaptive threshold to trigger the handover for \acp{uav}.

The inclusion of a new type of user in the network requires proper implementation from the initial planning stages of network deployment, up to its operation and optimisation. The challenges presented in this paper highlight the need for operators to take steps to prepare the network for the introduction of \ac{uav} users, otherwise the network may experience \ac{qos} issues for both air as well as ground users. 

In our future work, we intend to investigate the solutions to the issues presented in this paper through simulations, mathematical tools and real-world measurements. 
\section*{Acknowledgements}
This work was supported by a research grant from Science Foundation Ireland (SFI) and the National Natural Science Foundation Of China (NSFC) under the SFI-NSFC Partnership Programme Grant Number 17/NSFC/5224, as well as SFI Grants No. 16/SP/3804. It was also supported by the Commonwealth Cyber Initiative (CCI). The authors would like to thank Conor Duff and Gavin Lee from DenseAir for providing the Dublin \ac{uav} measurement dataset. 

\bibliographystyle{IEEEtran}
\bibliography{IEEEabrv,main}


\end{document}